# The dynamical evolution and the force model for asteroid (196256) 2003 EH1

T.Yu. Galushina, G.E. Sambarov*

*Tomsk State University, 634050, Tomsk, Russian Federation, phone +73822529776, e-mail detovelli@gmail.com

___

**Abstract**

This paper is devoted to the dynamics of asteroid (196256) 2003 EH1 that belongs to the Amor group. It is known that the asteroid 2003 EH1 is associated with one of the main annual meteor showers – the Quadrantids. In this work we analyze the influence of various perturbing factors on the asteroid motion. The perturbation's estimation was done by five different methods based on the nominal orbit evolution and the size of the initial confidence region. The most significant influences on the dynamical evolution of 2003 EH1 are gravitational forces from the Sun, major planets and the Moon, and the relativistic effects (RE) of the Sun. Of less importance are the Earth, the Sun and Jupiter oblateness; gravitational perturbations from Pallas, Ceres, Vesta and Pluto; and the RE of planets, the Moon, and Pluto.

The researches of chaoticity and evolution of asteroid 2003 EH1 were examined by integrating its motion equations along with 500 clones. The time interval (1000-4000 years) has been determined by integration precision estimation. We calculated the mean exponential growth factor of nearby orbits (MEGNO) and found that MEGNO < 2 only in the interval 1700-2300. After 2300 year the MEGNO parameter increases that indicates motion instability. It shows that the orbit may be considered as regular on the time interval of ±300 years from now, and as chaotic outside this interval. The reason, as we suppose, is frequent close approaches of the asteroid with Jupiter and the overlap of apsidal-nodal resonances.

___

## 1. Introduction

The motion simulation plays a special role in the study of the dynamical properties of Solar System objects. The choice of the approach for estimating an optimal model depends on many factors. The initial assumption concerning a force model for an investigated object is one of the most complicated and important aspects. Using the full model is not always justified. The accuracy of obtaining initial and current estimates depends on the correct construction of force model. Therefore, analysis of the force function should be the first step in the study of the asteroid orbital evolution.

In this study we investigate the structure of perturbing accelerations and dynamical evolution for asteroid (196256) 2003 EH1. The dynamical studies showed that the asteroid is associated with one of the main annual meteor showers – the Quadrantids (Jenniskens, 2003; Williams et al., 2004a; Porubcan & Kornos, 2005). The Quadrantids are among the most active meteor showers, reaching a peak activity on 3–4 January each year. The status of meteoroid stream modeling is still in its infancy. There is no correct unified model that describes the meteoroid streams, besides many models fail to explain even the most basic features of the observed meteor showers (Jenniskens et al., 2016). Each meteoroid streams need in individual approach. The age and formation mechanism of the core of the Quadrantid meteoroid stream was studied (Wiegert & Brown, 2005; Abedin et al., 2015). A current dust production from 2003 EH1 is too small (Kasuga & Jewitt, 2015) to supply the mass of the Quadrantids on interval 200–500 year ago from now. If 2003 EH1 is the source of the Quadrantids core then mass must be delivered episodically.

Since a Tisserand parameter with respect to Jupiter is $T_J$ = 2.066 the asteroid is on a Jupiter-family-comet type orbit but it shows no evidence of a cometary activity. The inactivity of 2003 EH1 points perhaps at an asteroidal nature. Jenniskens (2004) and Koten et al. (2006) concluded that 2003 EH1 is a dormant comet and that the Quadrantids get in a category between meteoroids of cometary and asteroid origin.

In this paper we study the relative importance of perturbing accelerations for asteroid 2003 EH1 to classify them as powerful or weak ones. The set of estimated perturbing accelerations contains: gravitational perturbations from all major planets, Pluto, the Moon, Ceres, Pallas, Vesta; the Earth, the Sun and Jupiter oblateness; the relativistic effects of the Sun, all major planets, Pluto, and the Moon. The force model with all listed perturbations will be called the full model.

The asteroid is briefly described in **Section 2**. To compare action of perturbing factors we used five indicators (Galushina et al., 2015), and the methods of their calculation are briefly described in **Section 3**, and also we present in this section methods for calculation of the mean exponential growth factor of nearby orbits (MEGNO) (Cincotta et al., 2003) and the algorithm for identifying the apsidal-nodal resonances (Bordovitsyna et al., 2014). The results and their discussion follow in **Section 4**.

All investigations have been carried out by software "IDA" developed in Tomsk State University to study the asteroid dynamics (Bykova et al., 2012a; Bykova et al., 2012b; Galushina, Bykova, 2008; Razdymakhina, 2011). The differential equations of motion have been integrated numerically by the Everhart method of 19th order with variable time step (Everhart, 1985).

## 2. Asteroid (196256) 2003 EH1

The asteroid (196256) 2003 EH1 is a near-Earth asteroid (NEA) of Amor type. It was discovered in March 2003 by the Lowell Observatory Near-Earth Object Search (LONEOS). The current orbit of the asteroid 2003 EH1 with the orbits of Mercury, Venus, the Earth, Mars and Jupiter is presented in **Fig. 1**. The black points on the asteroid orbit show the asteroid position in the observation moments. The 2003 EH1 has 99 observations in the database of IAU Minor Planet Center (MPC) on interval 4035.04 days and interval of observations is more than orbital period ($T = 5.52$ years). **Fig. 1** shows 2003 EH1 has regular distribution of observations on the orbit.

The orbital elements and physical parameters of 2003 EH1 are given in **Table 1**. It contains orbital elements (semi-major axis $a$, eccentricity $e$, inclination of orbit to ecliptic $i$, longitude of ascending node $\Omega$, argument of perihelion $\omega$ and mean anomaly $M$) and their uncertainties at epoch $t_0$ and $t_{JPL}$, absolute magnitude $H$, albedo $A$ and radius $R$ and information about number of observations $N$ and time interval of observations $\Delta t$ of 2003 EH1. The epoch $t_0$ is average of observation moments (2006 Nov, 11.0) and $t_{JPL}$ is 2016 July 31.0. We present the orbital elements obtained as the result of the orbit improvement with the full force model using software IDA (marked IDA) and from Jet Propulsion Laboratory (JPL) to compare. The orbital elements and their uncertainties are in good correspondence with each other. The physical parameters have been taken from the paper (Kasuga & Jewitt, 2015) and the observations have been received from the MPC. The asteroid has been observed over a span of 11 years, and so its orbital elements could be calculated with sufficient precision.

This asteroid is believed to be the parent body for the Quadrantids meteoroid stream, but its origin is not quite clear. The relationship between asteroid 2003 EH1 and the Quadrantids meteoroid stream has been studied previously by several authors (Jenniskens, 2004; Williams et al. 2004b; Wiegert & Brown 2005). To model meteoroid stream formation a precise parent body orbit is needed, so it is advisable to analyze influence of various perturbation factors on the asteroid motion and dynamical evolution. Therefore the presented analysis could be of interest for various groups of researchers.

## 3. Methods

*3.1. The investigation into the relative importance of perturbations on the asteroid orbit*

For the investigation of the perturbations' structure we apply special technique which consist from five various methods (Galushina et al., 2015). Let's describe these approaches. In all methods first step is determination of the initial time $t_0$ which is arithmetic mean of observations' moments. As known this moment is close to the best conditionality time (Syusina at el, 2012).

I. The improvement of the orbital elements is made using the full force model $F$. We calculate the orbital evolution using sub-models $F^*$ with one perturbation factor excluded. The maximal difference $\Delta r$ between position vectors calculated for $F$ and $F^*$ models reached on the integration time interval is used as the investigated indicator.

II. This time both improvement of the orbital elements and calculation of the orbital evolution are performed using models $F^*$. The investigated indicator is again $\Delta r$.

III. The improvement of the orbital elements is made using the full force model $F$. During the integration with model $F$ the maximum values of instantaneous accelerations $v = \ddot{x}_{max}$ for each perturbation are calculated and used as indicators.

IV. We make improvement of orbital elements using the model $F$ and also $F^*$ models, and calculate standard errors (O–C). The value $\psi = \sigma^2/\sigma_0^2$ is used as the indicator; here $\sigma_0$ is the mean square error for the model $F$ and $\sigma$ is the same, but for one of the $F^*$-models. The critical value is $\psi_{crit} = \chi^2_{N-K}/(N-K)$, where $\chi^2_{N-K}$ is Pearson's criterion for $(N-K)$ degrees of freedom, $N$ and $K$ are the number of observations and evaluated parameters respectively (99 and 6 in our case). If observation errors are random and have normal law distribution then $\psi$ is found in confidence interval $(0, \psi_{crit})$ with probability 0.999. If the observation errors have systematical part then $\psi > \psi_{crit}$.

V. We make improvement of orbital elements using the model $F$ and also $F^*$ models. The comparison of the model $F$ and models $F^*$ is made by means of the indicator $\varepsilon = |\hat{q}^* - \hat{q}|/|\bar{q} - \hat{q}|$. Here $\hat{q}$ is the vector of asteroid orbital parameters $\boldsymbol{q} = (q_1, ..., q_K)$, obtained by the least square method for the model $F$, and $\hat{q}^*$ is the same, but for one of the $F^*$-models; $\bar{q}$ is a point, which lies in the parametric space along direction $(\hat{q}^* - \hat{q})$ on the boundary surface of the asteroid's motion confidence region $\Phi_F$ (Syusina et al, 2013). If $\varepsilon > 1$ the asteroid motion turned out to be beyond $\Phi_F$, so the perturbing factor missing in the current $F^*$-model should be taken into account (i.e. without this factor the current $F^*$-model is too far from $F$).

Thus we can see that first, second and third methods are based on the analysis of orbital evolution, fourth method is founded on estimation of (O–C) residuals and fifth method relies on valuation of initial probability domain size. In the first three methods evolution of the asteroids orbits have been investigated on time interval equal 300 years. In the fourth and fifth method conclusions are made for moment $t_0$. In all cases planetary coordinates were taken from the Jet Propulsion Laboratory (JPL) Planetary Development Ephemeris – DE431. The relativistic effects are taken into account by adding Schwarzschild terms to the motion equation (Brumberg, 1972).

*3.2. Algorithm for calculating the MEGNO parameter*

The averaged mean exponential growth factor of nearby orbits (MEGNO) (Cincotta et al., 2003) is a time-weighted integral form of the Lyapunov characteristic number (LCN). Let's assume that the dynamic system is described by the following system of equations:

$$\frac{d}{dt}\mathbf{x}(t) = f(\mathbf{x}(t), \alpha), \mathbf{x} \in \mathbf{R}^{2n} \quad (1)$$

where $\mathbf{x}(t)$ is the six-dimensional system state vector and $\alpha$ is the vector of parameters of the forces model. Let $\phi(t) = \phi(t, x_0, t_0)$ be the solution of system (1) under the $(t_0, x_0)$ initial conditions. Relevant information about the flow in the vicinity of any orbit $\phi(t)$ is gained through its largest LCN $\lambda$ defined as

$$\lambda = \lim_{t \to \infty} \frac{1}{t} \ln \frac{\|\delta_\phi(t)\|}{\|\delta_\phi(t_0)\|}, \quad (2)$$

where $\delta_\phi(t)$ is the so-called tangent vector that measured the evolution of the initial infinitesimal difference $\delta_\phi(t_0) \equiv \delta_0$ between the $\phi(t)$ solution and a very close orbit. This evolution may be described with accuracy up to the first order infinitesimals by the following variational equation:

$$\dot{\delta}_\phi = \frac{d}{dt}\delta_\phi(t) = \mathbf{J}(\mathbf{f}(\phi(t)))\delta_\phi(t), \quad \mathbf{J}(\mathbf{f}(\phi(t))) = \frac{\partial \mathbf{f}}{\partial \mathbf{x}}(\phi(t)), \quad (3)$$

where $\mathbf{J}(\phi(t))$ is the Jacobian matrix of a system of differential equations. The fact that the LCN measures the "mean exponential rate of divergence of nearby orbits" is stated explicitly when recasting $\lambda$ in the integral form:

$$\lambda = \lim_{t \to \infty} \frac{1}{t} \int_0^t \frac{\dot{\delta}_\phi(s)}{\delta_\phi(s)} ds, \quad (4)$$

where $\delta_\phi = \|\boldsymbol{\delta}_\phi\|$, $\dot{\delta}_\phi(s) = \dot{\boldsymbol{\delta}}_\phi \cdot \boldsymbol{\delta}_\phi / \delta_\phi$.

The $Y_\phi(t)$ MEGNO parameter is introduced as a time-weighted integral form of LCN:

$$Y_\phi(t) = \frac{2}{t}\int_0^t \frac{\dot{\delta}_\phi(s)}{\delta_\phi(s)} s\, ds, \qquad (5)$$

the average $\bar{Y}_\phi(t)$ value is obtained in the following form:

$$\bar{Y}_\phi(t) = \frac{1}{t}\int_0^t Y_\phi(s)\, ds, \qquad (6)$$

The time evolution of $Y_\phi(t)$ and $\bar{Y}_\phi(t)$ values manifests certain features specific to different types of orbits. For example, it is known that $Y_\phi(t)$ for quasiperiodic (regular) orbits oscillates around 2, and $\bar{Y}_\phi(t)$ always tends towards 2. $\bar{Y}_\phi(t)$ for stable orbits of a harmonic oscillator type equals zero.

It is a good practice to substitute integral relations (5) and (6) with differential equations in numerical modeling problems and integrate, together with equations of motion (1) and variational equations (3), two more equations (Valk et. al., 2009):

$$\frac{d}{dt}y = \frac{\dot{\boldsymbol{\delta}}\cdot\boldsymbol{\delta}}{\boldsymbol{\delta}\cdot\boldsymbol{\delta}}t, \quad \frac{d}{dt}w = 2\frac{y}{t}, \qquad (7)$$

where $y$ and $w$ values are related to the MEGNO parameters in the following way:

$$Y(t) = 2y(t)/t, \quad \bar{Y}(t) = w(t)/t. \qquad (8)$$

*3.3. Algorithm for identifying the apsidal-nodal resonances*

As known (Murray and Dermott, 2009), the perturbing function describing the effect of the third body on an asteroid can be expressed in terms of orbit elements as follows (Bordovitsyna et. al., 2012):

$$\begin{aligned}
R = \frac{\mu'}{a'}\sum_{l=2}^{\infty}\alpha^l \sum_{\bar{m}=0}^{l}(-1)^{l-\bar{m}}\chi_{\bar{m}}\frac{(l-\bar{m})!}{(l+\bar{m})!} &\times \sum_{p,p'=0}^{l} F_{l\bar{m}p}(i)F_{l\bar{m}p}(i') \times \\
&\times \sum_{q,q'=-\infty}^{\infty} X_{l-2,p+q}^{l,l-2p}(e) X_{l-2,p'+q'}^{-l-1,l-2p'}(e') \times \\
&Cos[(l-2p'+q')\lambda' - (l-2p+q)\lambda \\
&-q'\varpi' + q\varpi + (\bar{m}-l-2p')\Omega' - (\bar{m}-l-2p)\Omega],
\end{aligned} \qquad (9)$$

where $\mu' = Gm'$ is the product of the gravitational constant and the mass of a disturbing body; $a, i, e, \Omega, \omega, M$ are the semi-major axis, eccentricity, longitude of the ascending node, argument of perihelion, mean anomaly of the asteroid orbit; $a', i', e', \Omega', \omega', M'$ are the same elements of the perturbing body's orbit; $\lambda = \varpi + M$ and $\lambda' = \varpi' + M'$ are the mean longitudes of the asteroid and the third body respectively; $\varpi = \Omega + \omega$ and $\varpi' = \Omega' + \omega'$ are the perihelion's longitudes of the asteroid and the disturbing body; $\alpha = a/a'$, $F...(i)$ is the inclination function; and $X_{...}(e)$ is the function of eccentricity, $\chi_0 = 2$, $\chi_{\bar{m}} = 2$ if $m \neq 0$.

The argument of the series expansion of the perturbing function (9) has the form

$$\underline{\psi} = (l - 2p' + q')M' - (l - 2p)\omega + (l - 2p')\omega' - \bar{m}(\Omega - \Omega'), \tag{10}$$

in the singly-averaged problem, and

$$\underline{\underline{\psi}} = (l - 2p')\omega' - (l - 2p)\omega - \bar{m}(\Omega - \Omega'), \tag{11}$$

in the doubly-averaged problem. In this case,

$$M' = M'_0 + \bar{n}'(t - t_0), \quad \omega' = \omega'_0 + \dot{\omega}'(t - t_0),$$

$$\Omega' = \Omega'_0 + \dot{\Omega}'(t - t_0), \tag{12}$$

$$\omega = \omega_0 + \dot{\omega}(t - t_0), \quad \Omega = \Omega_0 + \dot{\Omega}(t - t_0).$$

The condition of resonance occurrence can be represented as

$$\underline{\dot{\psi}} \approx 0, \quad \underline{\underline{\dot{\psi}}} \approx 0. \tag{13}$$

The secular accelerations in asteroid motion

$$\dot{\Omega} = \sum_{j=1}^{10} \dot{\Omega}_j,$$

$$\dot{\omega} = \sum_{j=1}^{10} \dot{\omega}_j. \tag{14}$$

are determined by the influence of the third body (Mercury ($j$=1), Venus ($j$=2), Earth ($j$=3), Mars ($j$=4), Jupiter ($j$=5), Saturn ($j$=6), Uranus ($j$=7), Neptune ($j$=8), Pluto ($j$=9) and Moon ($j$=10)):

$$\dot{\Omega}_j = -\frac{3}{16}\bar{n}m'_j\left(\frac{a}{a'}\right)^3 \frac{2+3e^2}{\sqrt{1-e^2}}(2 - 3\sin^2 i')\cos i,$$

$$\dot{\omega}_j = \frac{3}{16}\bar{n}m'_j\left(\frac{a}{a'}\right)^3 \frac{4 - 5\sin^2 i + e^2}{\sqrt{1-e^2}}(2 - 3\sin^2 i'). \tag{15}$$

Here $m'_j$ is the mass the third body. The inclination is significant for the value of resonance perturbations, which is why such resonances are called inclination-dependent.

A complete set of apsidal-nodal resonance relations with Jupiter is presented in **Table 2**.

## 4. Results and Discussion

*4.1. The investigation into the relative importance of perturbations on the asteroid orbit for (196256) 2003 EH1*

The values of indicators $\Delta r$, $\nu$, $\psi$ and $\varepsilon$ obtained by all five methods described in **Section 3.1** are shown in the **Fig. 2**. Horizontal lines on the figure show conditional division of perturbing factors on strong and weak ones (from left to right).

A comment regarding Method IV. In our case the mean square error for the full model $\sigma_0 = 0.522''$ and $\psi_{crit} = 1.345$. Therefore all perturbations with $\psi<1.345$ we may consider as weak factors (below the 'statistical' noise level). **Fig. 2** demonstrates that the perturbations from all the major planets, the Moon and the RE of the Sun are essential (Method V). Some difference exists in the perturbation factors classification. Nevertheless on the whole results of all methods are in the good agreement. The most important perturbations on the orbit of the asteroid are gravitational forces from the Sun, major planets and the Moon, the RE of the Sun. Of less importance are the Earth, the Sun and Jupiter oblateness, gravitational perturbations from Pallas, Ceres, Vesta and Pluto, the RE of planets. All these factors could be important for relatively short intervals of evolution – up to three hundred years.

*4.2. The investigation of the probabilistic orbital evolution of (196256) 2003 EH1*

The asteroid orbit has been improved by the least squares method. The results of orbital improvement are shown in **Table 3**, where $N$ is number of observations used in improving; $\Delta t$ is interval of observations in days and years; $t_0$ is the arithmetic mean of observations' moments which used as initial epoch; $\sigma$ is the mean square error of observations in arcseconds; $\sigma(X_0)$ is the mean square error of the least square estimation of vector $X_0$. The residuals (O-C) are presented in **Fig. 3** where $\delta$ is declination and $\alpha$ is right ascension. Four observations with (O-C)>1.263″ have been excluded by 3 sigma rule.

The observations distribute regular on the orbit, and a nonlinearity coefficient at time $t_0$ is less than the critical value of 0.1, that makes it possible to use the linear method of the initial confidence region construction. It is constructed in the six-dimensional phase space of coordinates and velocity components on the basis of the full covariance matrix in the form of an ellipsoid. The ellipsoid center is coordinates and velocities of the nominal orbit components resulting from improvement. We chose 500 normal distributed clones within initial confidence region.

Then we investigated the orbital evolution of each clone. The probabilistic orbital evolution study results are presented in **Fig. 4** and **5**. The **Fig. 4** shows evolution of semi-major axis $a$, eccentricity $e$, inclination $i$, longitude of ascending node $\Omega$, argument of perihelion $\omega$ and the parameter MEGNO. The evolution of clones is shown in gray, and the nominal orbit is marked out in black. The MEGNO parameter $\bar{Y} < 2$ only in the interval 1700-2300. After 2300 year the MEGNO parameter increases that says about motion instability. It indicates that the orbit may be considered as regular on the time interval of ±300 years from now, and as chaotic outside this interval. Thus initially close orbits diverge exponentially after ±300 years and **Fig. 4** and **5** show that the evolution of clones of (196256) 2003 EH1 is significantly different from the evolution of the nominal object outside the interval 1700-2300. The probability domain increases significantly after the year 2300 AD, thus it is impossible to predict reliably the asteroid motion.

*4.3. The influence of apsidal-nodal resonances on the orbital evolution of (196256) 2003 EH1*

The influence of Jupiter has significant actions on motion of the asteroid (**Fig. 2**). The reason is frequent close approaches of 2003 EH1 with this planet. **Fig. 5** shows the close encounters of nominal particle (black) and clones (gray) with Mercury, Venus, the Earth, Mars and Jupiter. As shown on **Fig. 5d** nominal orbit of 2003 EH1 hasn't close encounters with Mars, but its clones have close approaches. The asteroid is moving closer to Mercury, Venus and Earth than Mars, but the approaches for nominal object are not very strong – beyond the respective spheres of Hill.

As mentioned in the paper Jupiter is the dominant factor affecting this object and so we consider the apsidal-nodal resonances with it. It is well known that chaos arises where resonances overlap. The overlap of stable resonances of one spectral class can lead to chaotic motion of small bodies in the Solar System. In Chirikov (1979) the interaction of resonances is understood as the simultaneous influence of several resonances. In our case, we consider the full range of the apsidal-nodal resonances.

**Fig. 6** presents the evolution of apsidal-nodal resonance relationship and the critical arguments on interval 1000-4000 years. Numbers of resonant relations and their respective critical arguments given in the figures correspond to the numbering accepted in **Table 2**.

The character of evolution of the resonance relationships is such that they in most cases cross zero, i.e. sharp secular resonance occurs. The **Fig. 6** shows the lack of resonance $\psi_{11}$, while the apsidal-nodal resonances $\psi_9$, $\psi_{10}$, $\psi_{12}$, $\psi_{17}$, $\psi_{20}$ are present at small intervals, crossing zero twice. The eccentricity growth, characteristic for such perturbations, maybe has short-period character with overlapping of long-period apsidal-nodal resonances. The relationship

$$\sqrt{1-e^2}\cos i = const. \qquad (16)$$

is conserved at small time intervals (about 40 years). The close encounters with Jupiter are a cause of breaking of this relationship.

The close approaches with Jupiter and time variations in relationship (16) for asteroid 2003 EH1 are shown in **Fig. 7**. As may be seen, the value of this relationship varies within the limits of the second decimal digit over a 40 year time interval. If we consider the long-period evolution of the ratio (16), we can see that it is not conserved. A similar situation can be traced in the evolutionary picture of the resonance ratio $\psi_{20}$. Stable resonance configurations can be traced to short intervals of the study.

Based on behavior of the critical arguments which librate throughout the considered time interval, we note that apsidal-nodal resonances are stable. The movement of the critical arguments is a regular on all time interval. The amplitudes of the oscillations of the orbital parameters $\Omega$, $\omega$ are limited, when three main elements *a*, *i*, *e* change significantly over time. As a result, the long-term evolution of (196256) 2003 EH1 is characterized by large variations of the eccentricity and the inclination. The MEGNO parameter does not exceed 2 in the interval 1700-2300, but outside increases, which indicates appearance of chaoticity. Such interplay of resonances affects the long-term stability of orbit and the confidence regions.

## 5. Conclusions

The perturbation structure for an asteroid depends on its orbital elements, i.e. on its location and on its physical properties. In this paper we study effect of gravitational perturbations from all major planets, Pluto, the Moon, Ceres, Pallas, Vesta; the Earth, the Sun and Jupiter oblateness, the RE of the Sun, all major planets, Pluto and the Moon. All these factors could be important for relatively short intervals of evolution — up to three hundred years. To compare the perturbing factors effects we use five indicators, which allow us to classify perturbations as strong and weak ones. The most significant influences on the dynamical evolution of 2003 EH1 are gravitational forces from the Sun, major planets and the Moon, the relativistic effects of the Sun.

The presented results of studying the orbital evolution of (196256) 2003 EH1 show that the orbit of this object with the chosen initial motion parameters changes significantly. These alterations are attributed to close encounters with Jupiter and the action and overlapping of various apsidal-nodal resonances. The presence of these resonances leads to a significant increase in eccentricity, and their overlapping results in chaotization of the motion of asteroid.

The MEGNO parameter $\bar{Y} < 2$ only in the interval 1700-2300. After 2300 year the MEGNO parameter increases that indicates motion instability. It shows that the orbit may be considered as regular on the time interval of ±300 years from now, and as chaotic outside this interval.

**Acknowledgements**

One of the authors (SGE) would like to acknowledge the Organizers of the Meteoroids 2016 conference. This work was supported by the D.I. Mendeleev Scientific Foundation of Tomsk State University (project 8.1.54.2015).

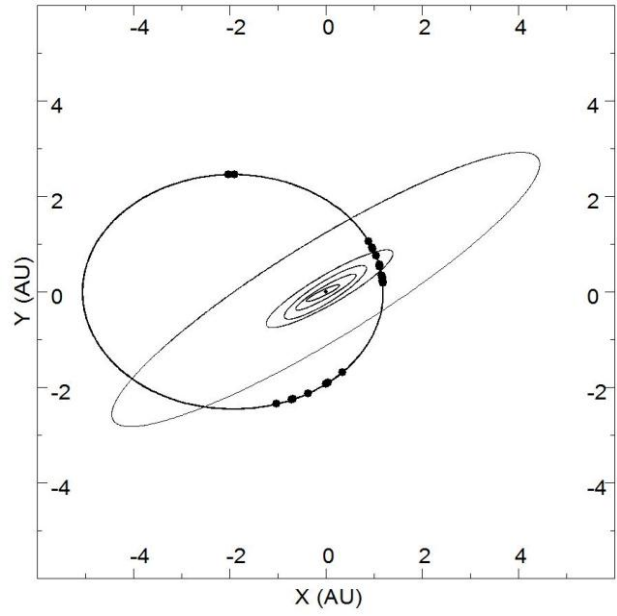

**Fig 1.** The distribution of observed positions of asteroid (196256) 2003 EH1 on projection on its orbital plane.

Table 1

Orbit elements, physical parameters and information about observations of (196256) 2003 EH1

| $t_0$ or $t_{JPL}$ | 2006 November 11.0 | | 2016 July 31.0 | | | |
|---|---|---|---|---|---|---|
| | IDA | | JPL | | IDA | |
| | Value | Uncertainty | Value | Uncertainty | Value | Uncertainty |
| $a$ (au) | 3.1263813830 | $2.3535 \cdot 10^{-6}$ | 3.1229356658 | $4.2279 \cdot 10^{-8}$ | 3.1229356404 | $2.8671 \cdot 10^{-6}$ |
| $e$ | 0.6183959485 | $6.2905 \cdot 10^{-7}$ | 0.6188099785 | $1.9319 \cdot 10^{-7}$ | 0.6188098863 | $6.4383 \cdot 10^{-7}$ |
| $i$ (deg) | 70.7763019730 | $1.5956 \cdot 10^{-7}$ | 70.8550736258 | $4.7303 \cdot 10^{-5}$ | 70.8551010631 | $3.4687 \cdot 10^{-7}$ |
| $\Omega$ (deg) | 282.9576698948 | $5.9455 \cdot 10^{-7}$ | 282.9771359501 | $5.2852 \cdot 10^{-5}$ | 282.9771526126 | $2.1269 \cdot 10^{-7}$ |
| $\omega$ (deg) | 171.3340280548 | $8.6951 \cdot 10^{-7}$ | 171.3717906942 | $7.7776 \cdot 10^{-5}$ | 171.3718992459 | $3.6306 \cdot 10^{-7}$ |
| $M$ (deg) | 241.8394906380 | $1.3432 \cdot 10^{-6}$ | 155.5947044838 | $2.4561 \cdot 10^{-5}$ | 155.5946832340 | $8.0121 \cdot 10^{-7}$ |
| $H$ (mag) | 16.2 | | | | | |
| $A$ | 0.04 | | | | | |
| $R$ (km) | 2 | | | | | |
| $N$ | 99 | | | | | |
| $\Delta t$ (days) | 4035.04 | | | | | |

Table 2

The apsidal-nodal resonances relations

| № | The resonance relation | № | The resonance relation | № | The resonance relation |
|---|---|---|---|---|---|
| 1 | $(\dot{\Omega} - \dot{\Omega}'_{Jup}) + \dot{\omega} - \dot{\omega}'_{Jup}$ | 8 | $(\dot{\Omega} - \dot{\Omega}'_{Jup}) - 2\dot{\omega} - 2\dot{\omega}'_{Jup}$ | 15 | $(\dot{\Omega} - \dot{\Omega}'_{Jup}) - 2\dot{\omega}'_{Jup}$ |
| 2 | $(\dot{\Omega} - \dot{\Omega}'_{Jup}) - \dot{\omega} + \dot{\omega}'_{Jup}$ | 9 | $(\dot{\Omega} - \dot{\Omega}'_{Jup}) + 2\dot{\omega}$ | 16 | $(\dot{\Omega} - \dot{\Omega}'_{Jup}) + 2\dot{\omega}'_{Jup}$ |
| 3 | $(\dot{\Omega} - \dot{\Omega}'_{Jup}) + 2\dot{\omega} - 2\dot{\omega}'_{Jup}$ | 10 | $(\dot{\Omega} - \dot{\Omega}'_{Jup}) - 2\dot{\omega}$ | 17 | $(\dot{\Omega} - \dot{\Omega}'_{Jup})$ |
| 4 | $(\dot{\Omega} - \dot{\Omega}'_{Jup}) - 2\dot{\omega} + 2\dot{\omega}'_{Jup}$ | 11 | $(\dot{\Omega} - \dot{\Omega}'_{Jup}) + \dot{\omega}$ | 18 | $\dot{\omega} - \dot{\omega}'_{Jup}$ |
| 5 | $(\dot{\Omega} - \dot{\Omega}'_{Jup}) + \dot{\omega} + \dot{\omega}'_{Jup}$ | 12 | $(\dot{\Omega} - \dot{\Omega}'_{Jup}) - \dot{\omega}$ | 19 | $\dot{\omega} + \dot{\omega}'_{Jup}$ |
| 6 | $(\dot{\Omega} - \dot{\Omega}'_{Jup}) - \dot{\omega} - \dot{\omega}'_{Jup}$ | 13 | $(\dot{\Omega} - \dot{\Omega}'_{Jup}) + \dot{\omega}'_{Jup}$ | 20 | $\dot{\omega}$ |
| 7 | $(\dot{\Omega} - \dot{\Omega}'_{Jup}) + 2\dot{\omega} + 2\dot{\omega}'_{Jup}$ | 14 | $(\dot{\Omega} - \dot{\Omega}'_{Jup}) - \dot{\omega}'_{Jup}$ | | |

## Method 1

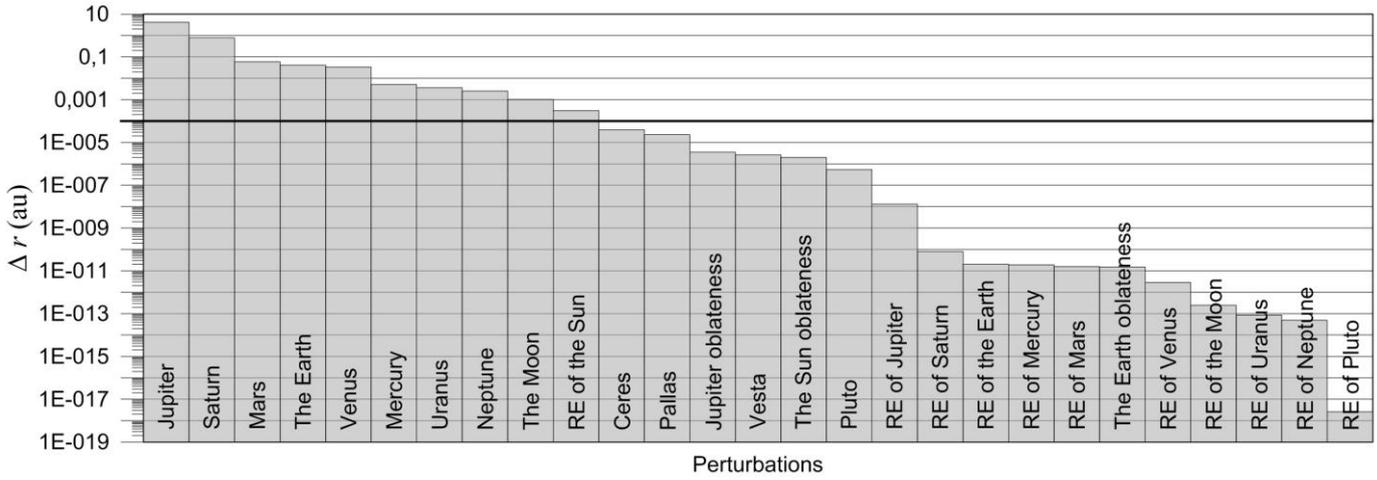

## Method 2

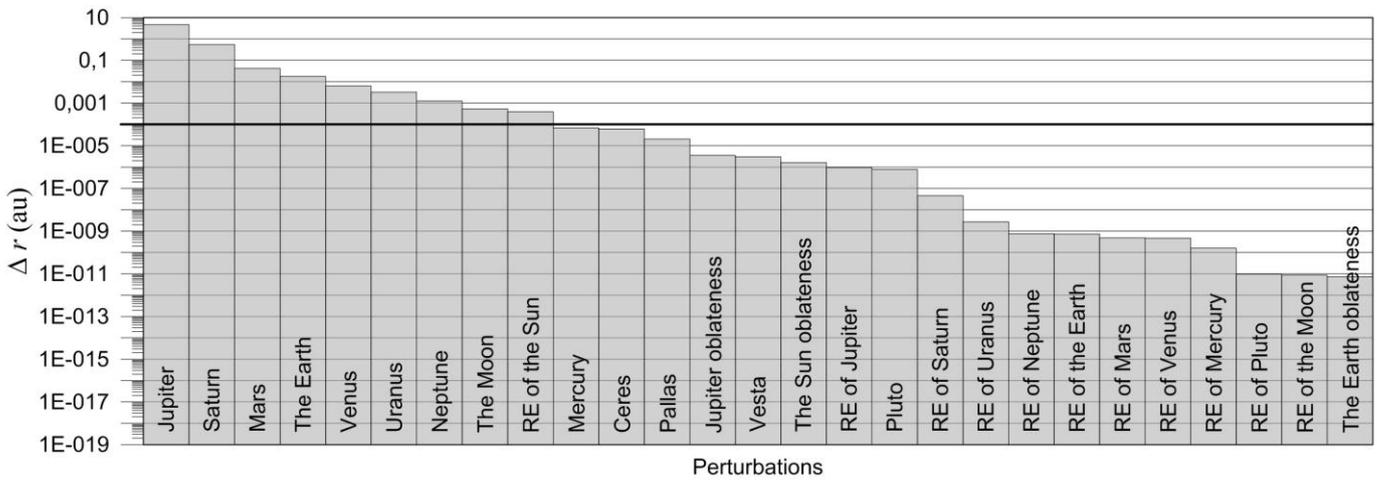

## Method 3

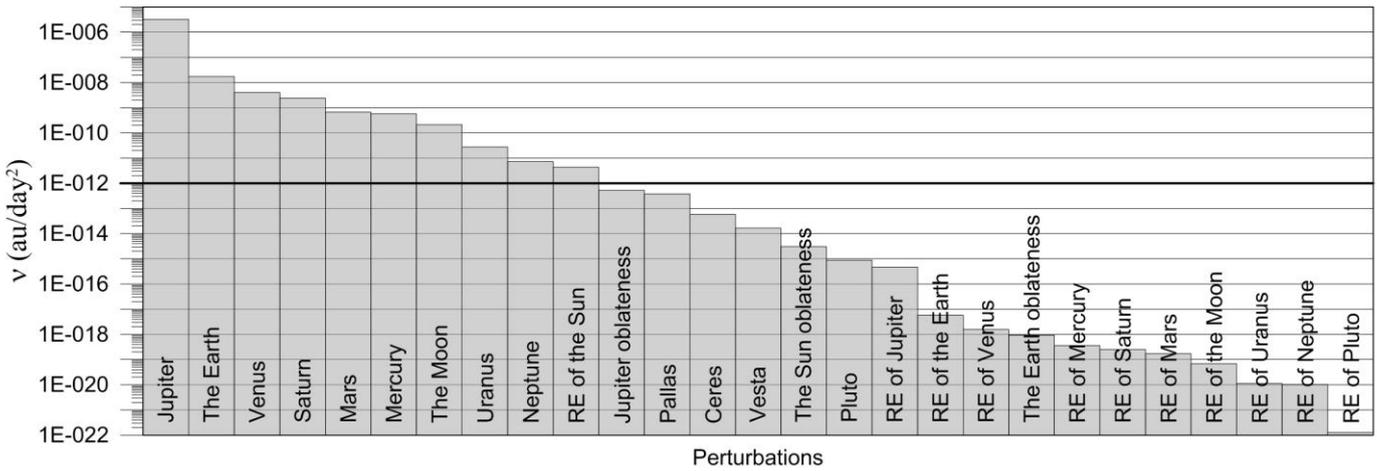

Method 4

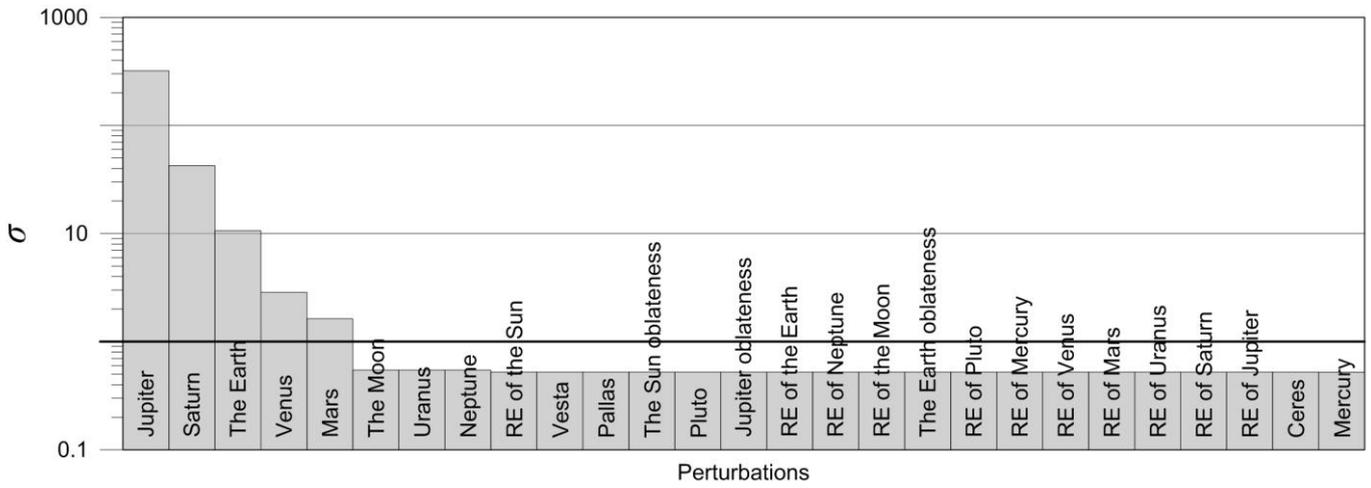

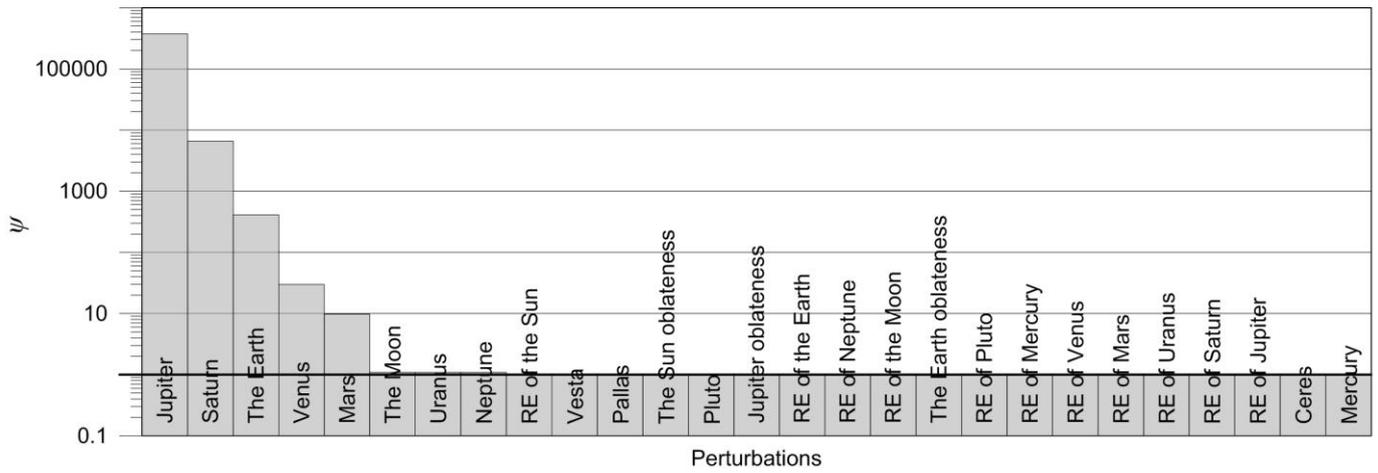

Method 5

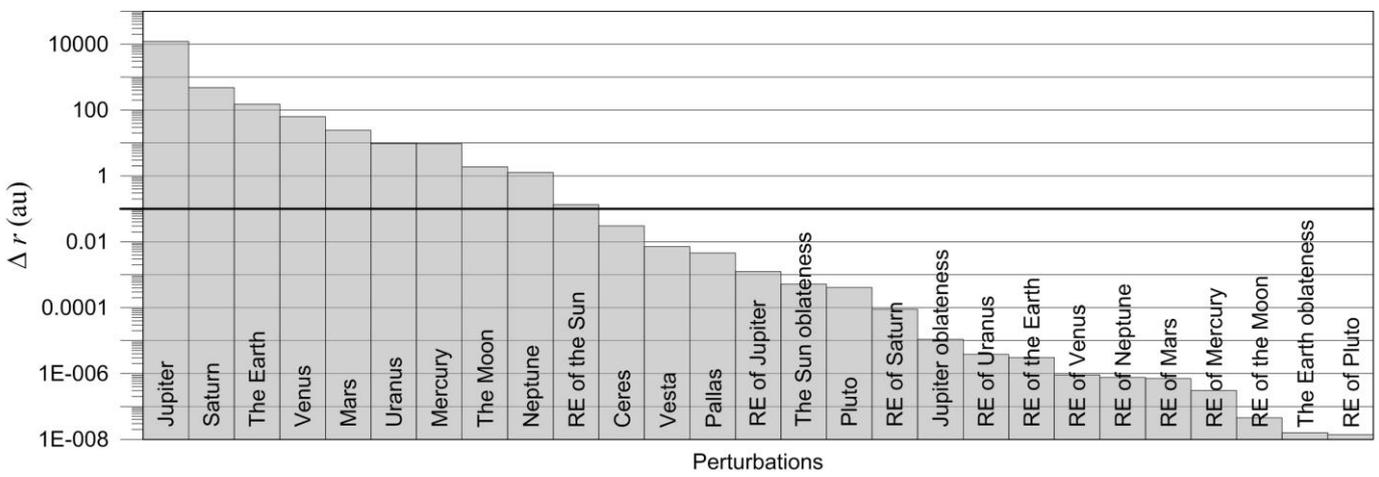

**Fig 2.** The results of perturbation structure study obtained for asteroid 2003EH1.

Table 3

Data about the observations and orbit improvement results for the asteroid 2003EH1

| N | 95 |
|---|---|
| $\Delta t$, days (years) | 4035 (2003–2014) |
| $t_0$ | 11.11.2006 |
| $\sigma$, " | 0.421 |
| $\sigma(X_0)$, au | $3\cdot10^{-6}$ |

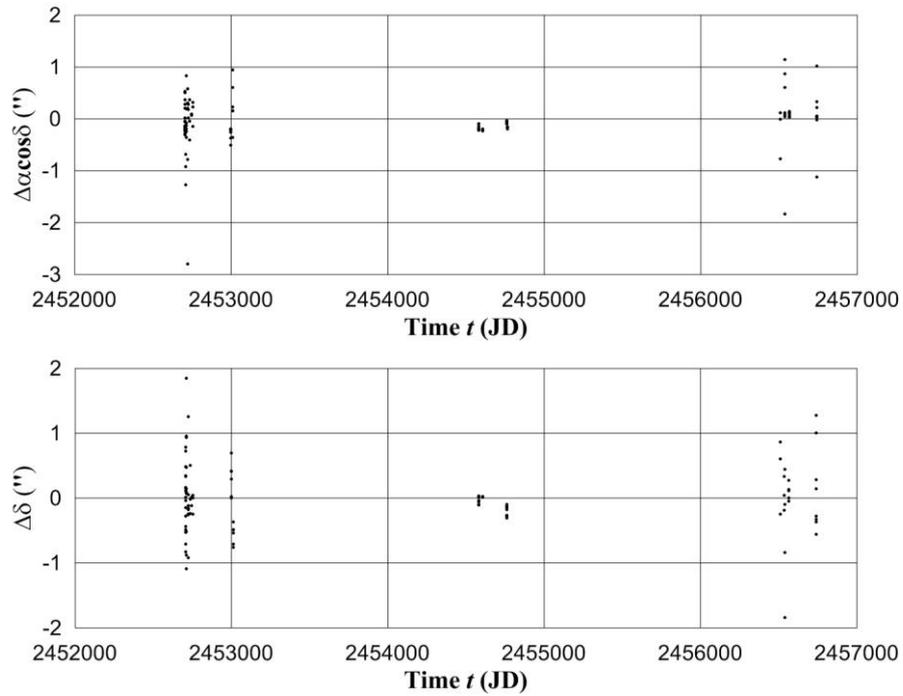

**Fig. 3.** The residuals (O-C) of (196256) 2003 EH1 observations

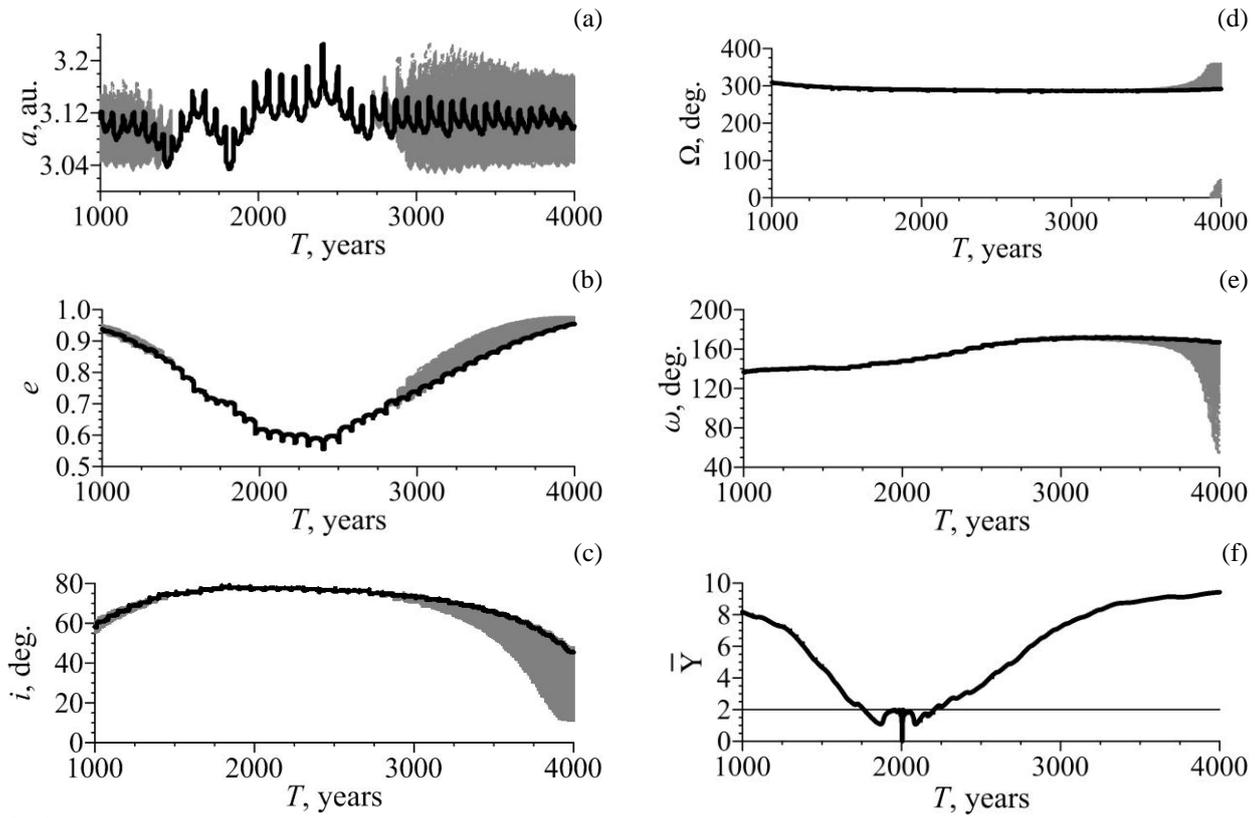

**Fig 4.** The results of the probability orbital evolution investigation for 2003 EH1: evolution of the semi-major axis(a), the eccentricity(b), the inclination of the orbital plane to equator (c), the argument of perihelion(d), the longitude of the ascending node(e) and MEGNO parameters(f) over a time interval of 3000 years (the nominal particle is black points and 500 clones are gray).

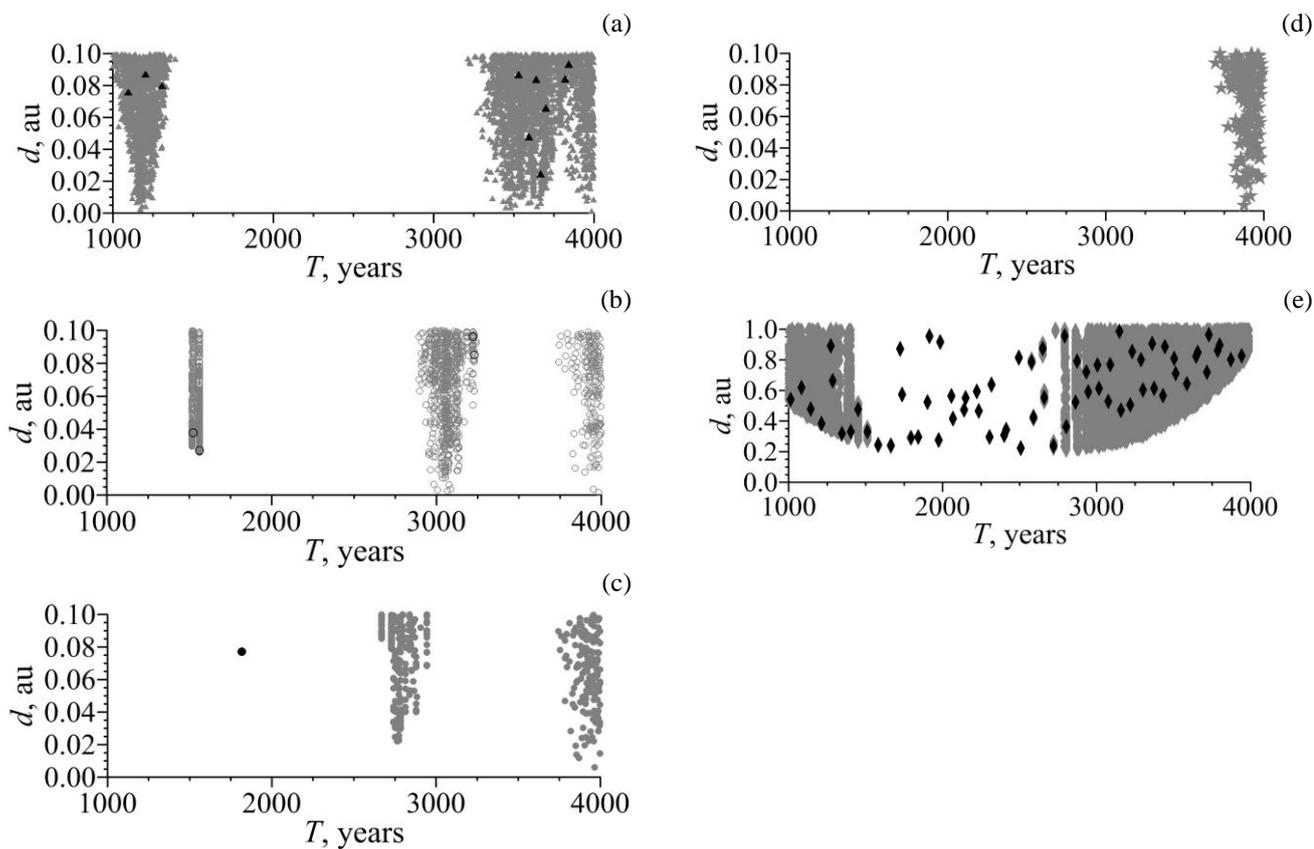

**Fig 5.** The asteroid 2003EH1: close approach of nominal particle (black) and 500 clones (gray) with Mercury (a), Venus (b), the Earth (c), Mars (d) and Jupiter (e), *d* is a distance between an object and planet.

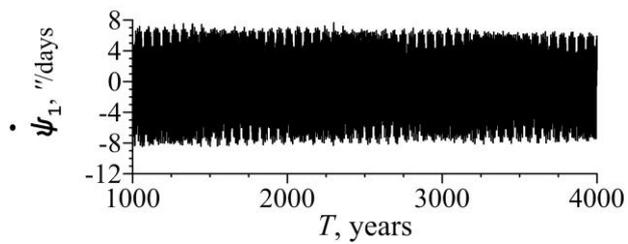
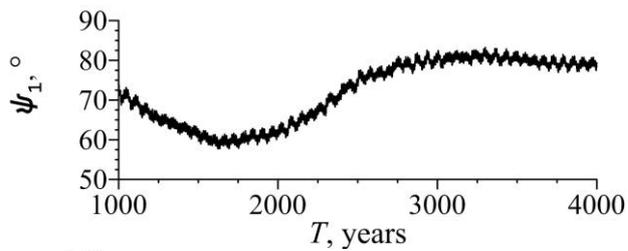
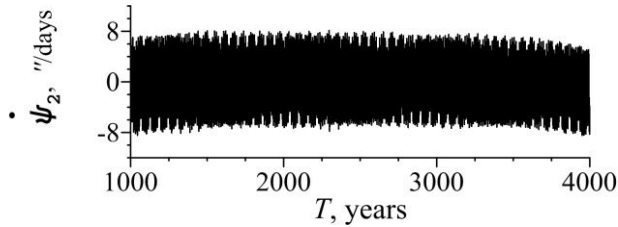
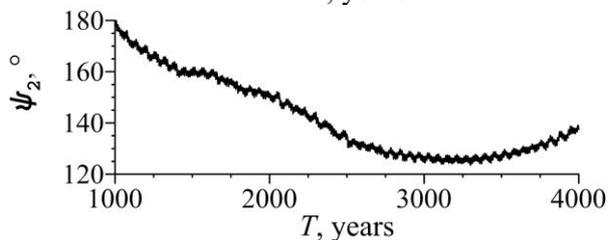
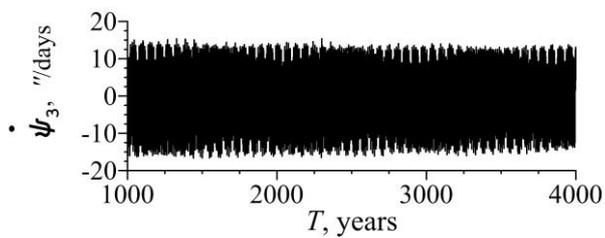
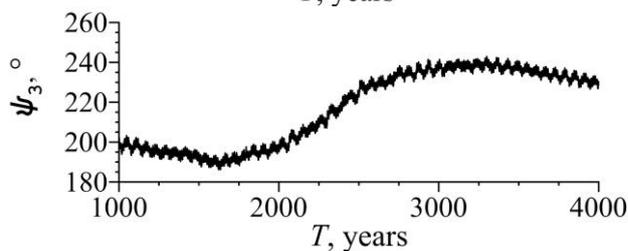
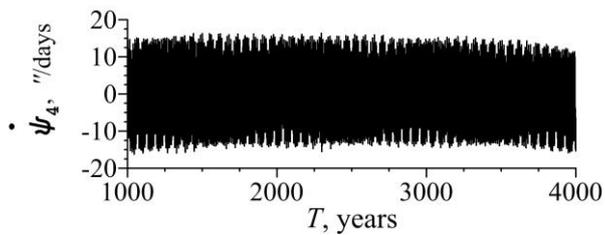
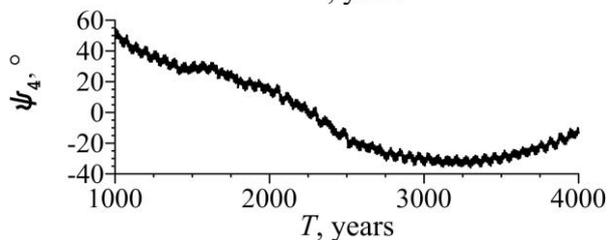
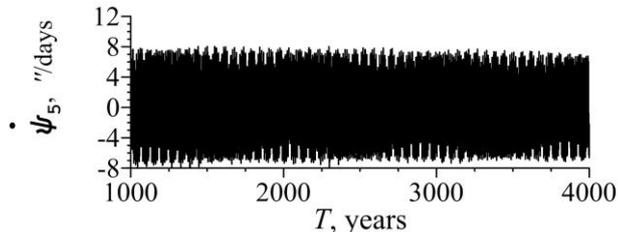
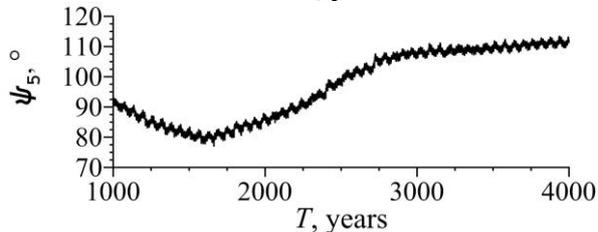
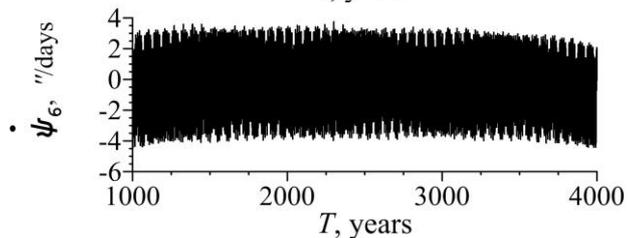
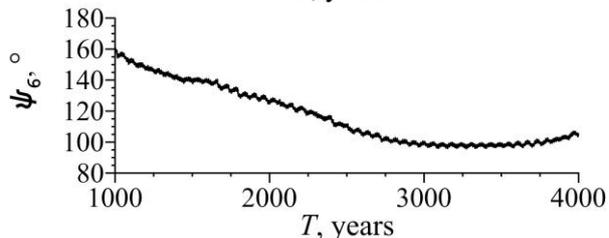
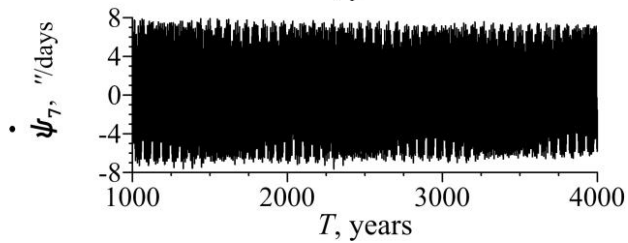
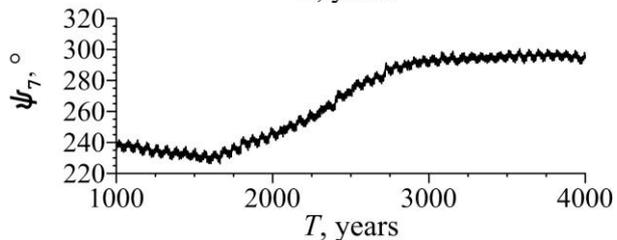

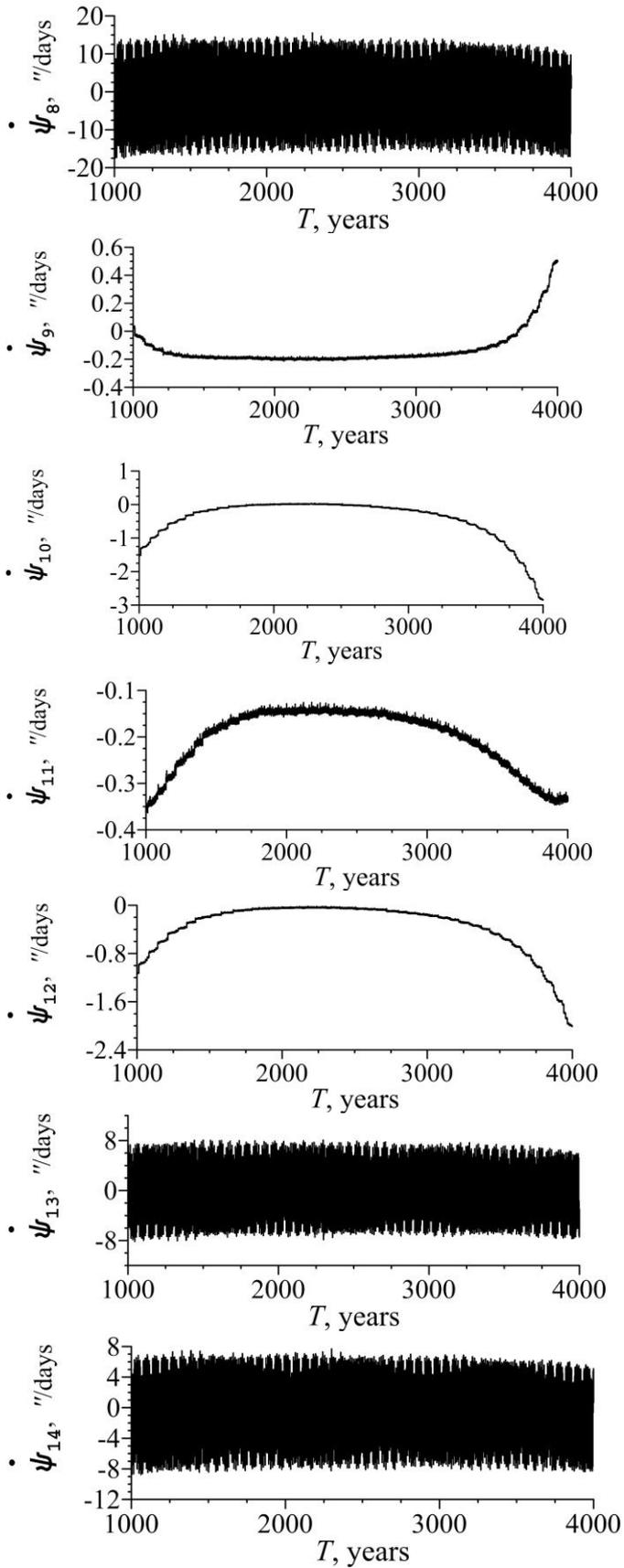
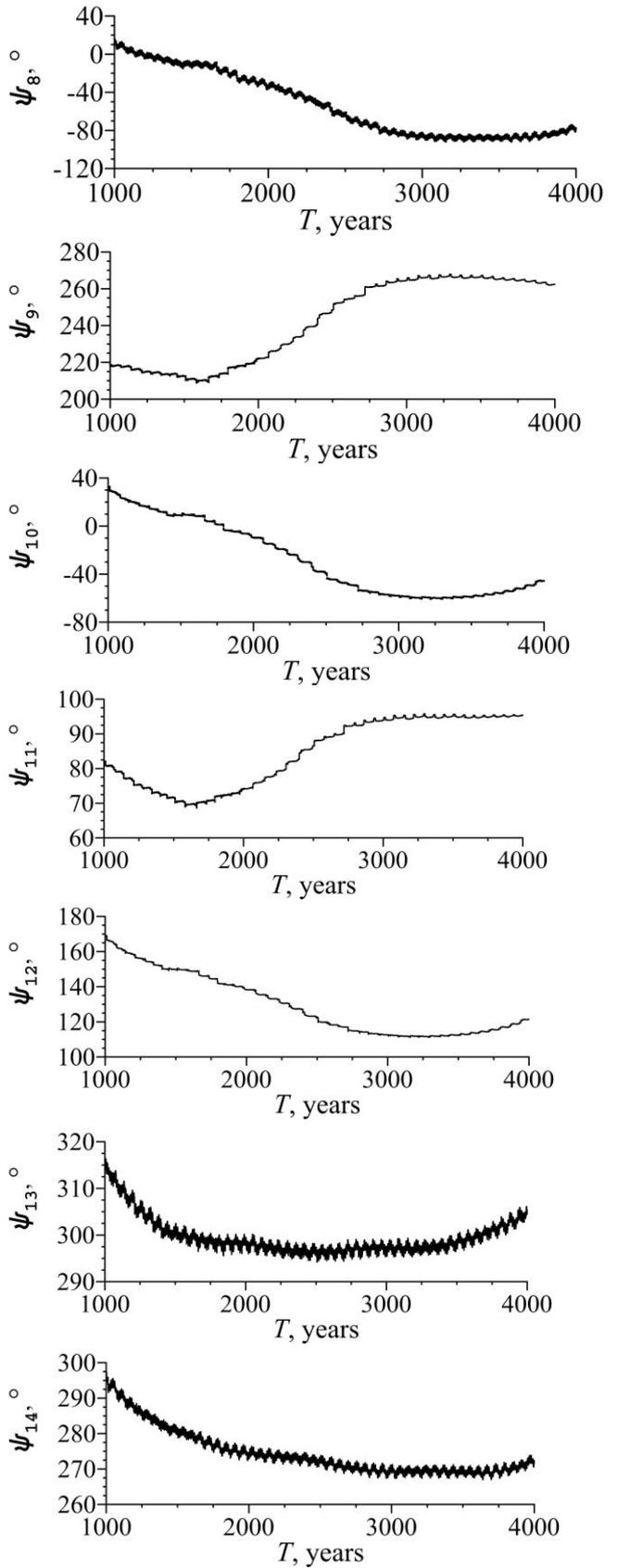

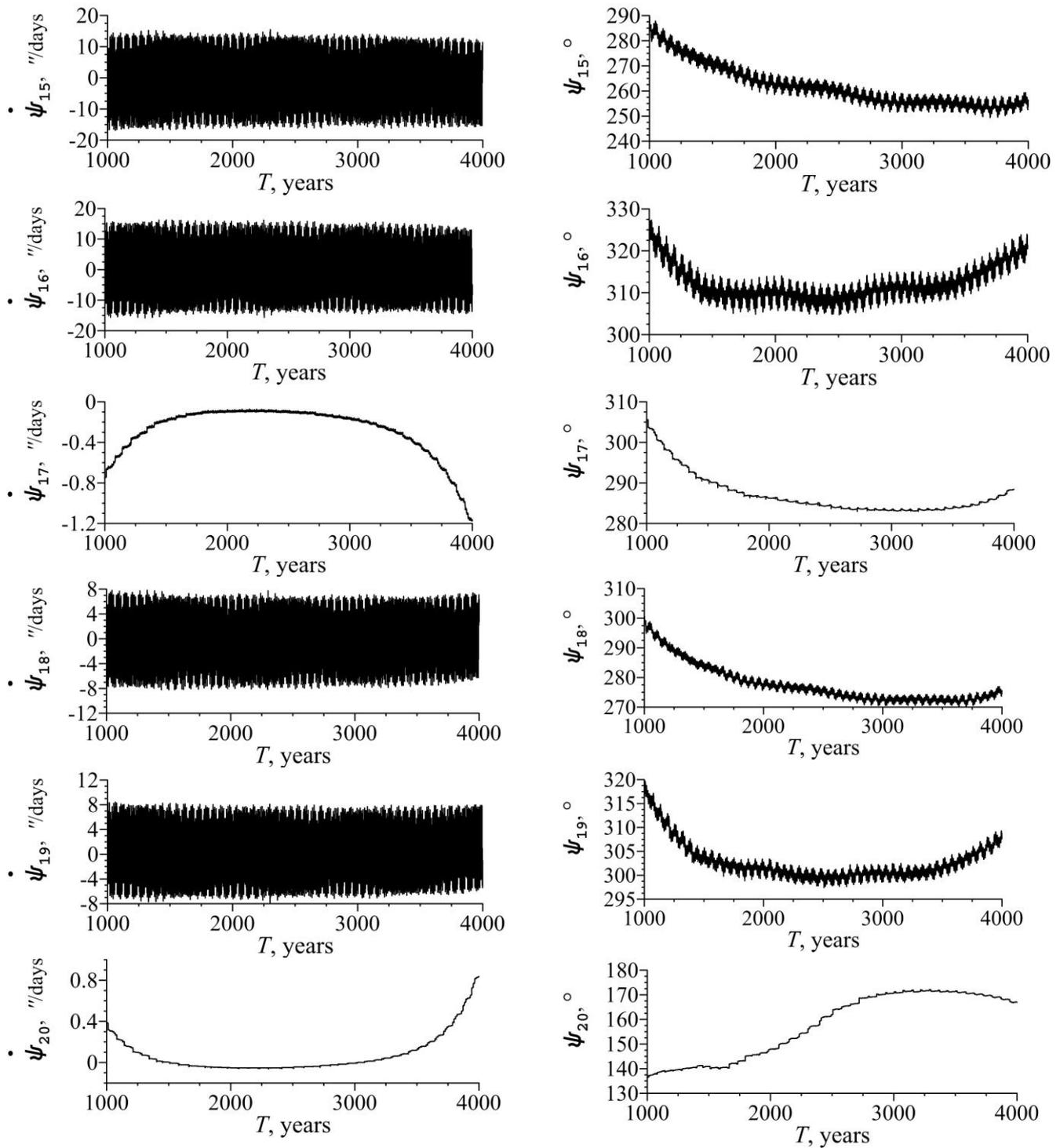

**Fig 6.** Evolution of the resonance relations and the critical arguments for asteroid (196256) 2003 EH1.

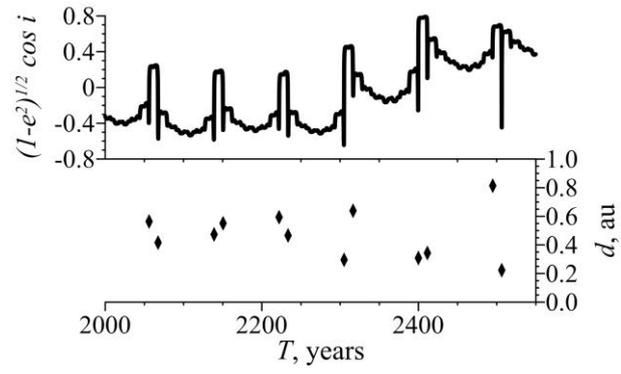

**Fig 7.** Time variations in relationship (16) (top). Close approach of asteroid 2003EH1 with Jupiter, *d* is distance between an object and Jupiter (bottom).